\documentclass[twocolumn,preprintnumbers,amsmath,amssymb,prl]{revtex4}

\newcommand\norm[1]{\left\lVert#1\right\rVert}

\usepackage{graphicx}
\usepackage{dcolumn}
\usepackage{bm}
\usepackage{amsmath}
\usepackage{float}
\usepackage{siunitx}
\usepackage{textcomp}
\usepackage{comment}
\usepackage{xcolor}
\usepackage{bbold}
\usepackage{bbm}
\usepackage{siunitx}

\begin{document}


\title{Monte-Carlo wavefunction approach\\for the spin dynamics of recombining radicals}
\author{Robert H. Keens}
\author{Daniel R. Kattnig}
 \email{D.R.Kattnig@exeter.ac.uk} 
\affiliation{%
Living Systems Institute and Department of Physics, University of Exeter, Stocker Road, Exeter, Devon, EX4 4QD, United Kingdom
 }%

\begin{abstract}

We adapt the Monte-Carlo wavefunction (MCWF) approach to treat the open-system spin dynamics of radical pairs subject to spin-selective recombination reactions. For these systems, non-Lindbladian master equations are widely employed, which account for recombination via the non trace-preserving Haberkorn superoperator in combination with reaction-dependent exchange and singlet-triplet dephasing terms. We show that this type of master equation can be accommodated in the MCWF approach, by introducing a second type of quantum jump that accounts for the reaction simply by suitably terminating the propagation. In this way, we are able to evaluate approximate solutions to the time-dependent radical pair survival probability for systems that have been considered untreatable with the master equation approach until now. We explicate the suggested approach with calculations for radical pair reactions that have been suggested to be relevant for the quantum compass of birds and related phenomena.

\end{abstract} 
 
\maketitle

\section{Introduction}

Interest in spin dynamics to explain biophysical phenomena has grown markedly in recent years, with a particular focus on magnetoreception \cite{ritz2000model, schulten1978biomagnetic, hore2016radical}, which is a cornerstone of the emerging field of quantum biology \cite{lambert2013quantum, mcfadden2016life}. For these systems, magnetosensitivity emerges from the reaction dynamics of radical systems, subject to coherent evolution, predominantly under Zeeman and hyperfine interactions, and spin-selective recombination \cite{steiner1989magnetic, brocklehurst2002magnetic, salikhov1984spin, hore2016radical, mclauchlan1991invited}. The quantum dynamics of the associated spin-degrees of freedom generally have to be considered in the presence of decoherence, relaxation and chemical reaction processes to give a realistic account of magneto-biological effects \cite{kattnig2017radical, kattnig2016electron, hiscock2016quantum, fay2018spin, lewis2016efficient}. Further, it has been shown that toy models are often insufficient to adequately describe these systems, as peculiar effects can arise from many-spin interactions, e.g. the emergence of spiky features in models of the radical-pair based compass \cite{hiscock2016quantum}, or often neglected interactions, e.g. the electron-electron dipolar coupling \cite{babcock2020electron, kattnig2016electron}. In fact, the study of simple models might also mislead, as effects might emerge that do not generalize to radical systems under realistic conditions, e.g. with many coupled nuclear spins or the open quantum system-setting. For example, long-lived entanglement has been attributed a role in radical-pair magnetoreception, but simulations with many coupled nuclei suggest a swift decay and no essential function \cite{master_students}. While some studies have started to approach more comprehensive systems/interaction levels, this typically applies to one interaction/phenomenon at a time and none can be said to have realized a truly comprehensive theoretical description of a realistic, biologically relevant radical pair system. Furthermore, new magnetic field effects have been predicted to result from the electron-electron dipolar interaction in systems of more than two radicals, which demonstrates the demand to simulate ever larger systems \cite{keens2018magnetosensitivity,sampson2019magnetosensitivity,babcock2020electron}.

The challenge with spin dynamics in general is that the size of Hilbert space grows exponentially with the number of considered spins. This is particularly stringent for the dynamics of open quantum systems which do not preserve pure states and thus mandate a formulation in Liouville space, which scales quadratically in Hilbert space dimension \cite{breuer2002theory}. This often leaves theoreticians with the choice between modelling small, but often biologically irrelevant processes, or systems of moderate size that are simplified to an extent that they are unrealistic.

For typical open quantum systems, i.e. master equations of the standard Lindblad form, a significant step toward solving (more) realistic problems came with the introduction of the Monte Carlo Wave Function (MCWF) approach \cite{molmer1993monte, kornyik2019monte, plenio1998quantum}. The MCWF method, which is also known as the quantum jump method, substitutes the evolution of the density matrix with an ensemble average over individual quantum trajectories of wave functions evolved under a pseudo-Hamiltonian and subject to quantum jumps, i.e.\ discontinuous modifications of the wavefunction, which together account for the openness of the system. Unfortunately, the spin dynamics of radical pairs does not conform to standard Lindblad form if asymmetric (i.e.\ different in the singlet and triplet configurations) reactivity is included by the Haberkorn approach and its recent additions \cite{haberkorn1976density, jones2010spin, jones2011reaction, fay2018spin}. This means that the MCWF approach cannot be robustly applied to solve large spin dynamics problems.

Spin-selective recombination processes are an integral part of the spin dynamics of radical systems.\cite{steiner1989magnetic, brocklehurst2002magnetic, salikhov1984spin, mclauchlan1991invited} Various ways to include this aspect in the master equations have been discussed in the recent literature  \cite{jones2010spin, jones2011reaction, fay2018spin, kominis2009quantum, kominis2015radical}. The traditional approach, due to Haberkorn \cite{haberkorn1976density}, suggests that singlet (rate constant $k_S$) and triplet ($k_T$) recombination are to be accounted for by the superoperator
\begin{equation}
\hat{\hat{K}}\rho = -\frac{k_S}{2} \left\{P_S, \rho \right\} - \frac{k_T}{2} \left\{P_T, \rho\right\},
\label{K}
\end{equation}
where $P_{S, T}$ are the singlet and triplet projection operators, respectively. This gives rise to a non-trace preserving equation of motion of the (concentration-weighted) density operator, for which the trace of $\rho$ gives the survival probability of the radical (pair) systems. The form of eq.\ \ref{K} has been debated \cite{purtov2010theory, kominis2009quantum, jones2011reaction, jones2010spin, kominis2015radical, tsampourakis2015quantum, kominis2016reply, jeschke2016comment} and confirmed \cite{tiersch2012open, jones2011reply, ivanov2010consistent}. Recently, Fay et al.\ have suggested a series of quantum master equations of the recombining radical pair, which they derive from a microscopic description of the electron transfer reaction using the Nakajima–Zwanzig projector approach \cite{fay2018spin}. To second order in the electronic coupling of radical pair and product states (i.e.\ in the nonadiabatic limit), the authors recover the Haberkorn master equation augmented  with an additional reactive exchange coupling term (which was also discussed in \cite{kattnig2011magnetic}). In the fourth order in the coupling (i.e.\ for more adiabatic reactions) an additional singlet-triplet dephasing (S/T-dephasing) term \cite{shushin1991effect} of the form
\begin{equation}
\hat{\hat{K'}}\rho = -k_{ST}\left(P_S\rho P_T + P_T\rho P_S\right)
\label{Kprime}
\end{equation}
appears, the rate $k_{ST}$ of which depends on the details of the system. It is noteworthy that additional S/T-dephasing earlier emerged from Kominis' theory \cite{kominis2015radical, kominis2009quantum, kritsotakis2014retrodictive}. The quantum measurement approach to radical pair recombination by Jones and Hore can also be interpreted as a result of the combined effect of $\hat{\hat{K}}$ and $\hat{\hat{K'}}$, in this case with the particular choice of $k_{ST} = (k_{S} + k_{T} )/2$ \cite{jones2010spin, jones2011reaction}. This suggests a central role of the Haberkorn approach augmented by additional S/T dephasing for the spin dynamics of radical systems, which is in fact reflected in the widespread use of this combination for the modelling of experimental data \cite{hoang2018magnetic, shushin1991effect}.

If $\hat{\hat{K}}$ from eq.\ \ref{K} is the only non-coherent term in the master equation, the spin dynamics of the system can still be evaluated comparably cheaply, as $\rho$(t) can be constructed from an approach based on the propagation of wavefunctions under a non-Hermitian, effective Hamiltonian (of the form given by eq.\ \ref{Heff} below) \cite{lewis2016efficient}. Thus the simulation process can be handled entirely in the comparably small Hilbert space. Singlet-triplet dephasing as given by $\hat{\hat{K'}}$ in eq.\ \ref{Kprime}, does not allow for this quasi-pure state evolution approach. Using $P_S = 1 - P_T$, we can however rewrite eq.\ \ref{Kprime} in many equivalent ways, one of which is
\begin{equation}
\label{NewKp}
\hat{\hat{K'}}\rho = 2k_{ST} \left(P_S\rho P_S - \frac{1}{2} \left\{P^{\dagger}_S P_S, \rho \right\}\right).
\end{equation}
As eq. \ref{NewKp} is of the form of a Lindblad dissipator, the dynamics it induces can in principle be accounted for by the MCWF approach \cite{molmer1993monte, kornyik2019monte, plenio1998quantum}. However, for this to provide a feasible approach to the spin dynamics of radical systems, we require a means to extend the MCWF approach to also include the non-unitary contributions associated with the asymmetric recombination of radical pairs as described by eq.\ \ref{K}. This is the aim of this contribution.

Note that quantum trajectories have previously been suggested for the modelling of radical pair dynamics \cite{kritsotakis2014retrodictive}. However, it has been argued that Haberkorn's theory cannot be cast in terms of quantum trajectories \cite{tsampourakis2015quantum}. Our extension to the MCWF approach allows it to be applied to non-Lindbladian master equations resulting from terms of the form of eq. \ref{K}. This broadens the applicability of MCWF to the spin dynamics of radical systems subject to spin-selective reactivity. The approach accommodates singlet-triplet dephasing, and thus applies to the description of chemical radical pair reactivity beyond the Haberkorn approach. It is also applicable to models that apply the
Haberkorn reaction operator (or its extensions) to the dynamics of open quantum systems of Lindblad form, e.g. random field relaxation or Redfield type relaxation superoperators of arbitrary genesis \cite{kattnig2017radical, worster2016spin, kattnig2016electron, kattnig2016electron2}. In the limit of infinite samples, the MCWF model agrees exactly with the direct integration of the master equation. We show that the approach allows us to obtain estimates of the magnetic field effects of large spin systems, which we consider (currently) intractable by the direct integration method. 

This manuscript is structured as follows: first, we present a derivation of the extended MCWF approach and explain the unique steps taken to make it applicable to the non-Lindbladian recombination term, as shown in eq.\ \ref{K}. Then, we present some results obtained with this approach, and demonstrate their equivalence to those obtained with the direct integration of the master equation, showing also the comparison between the efficiency and numerical error of both methods. Finally, we suggest applications for the approach, and ways to further increase its capability.

\section{Derivation}

The MCWF method aims to reconstruct the equation of motion of the (spin) density operator from the ensemble average of stochastic quantum trajectories of state vectors \cite{molmer1993monte, kornyik2019monte, plenio1998quantum}. The approach predominantly provides a computational tool, the efficiency of which rests on the reduction in dimensionality (associated with treatment of Hilbert space state vectors instead of density operators) and the fact that often a relatively small (relative to the Hilbert space dimension) number of samples is sufficient to adequately reconstruct the observables of interest. While individual MCWF trajectories do not necessarily convey reality, they have occasionally been interpreted to do just that, i.e. to reflect the behaviour of single realisations of quantum systems.
We consider the master equation:
\begin{equation}
    \begin{aligned}
    &\frac{d \rho}{dt} = -i [H, \rho] + \hat{\hat{D}} \rho + \hat{\hat{K}} \rho \\
    & \\
    & =  -i [H, \rho] + \sum^M_m\left( J_m \rho J^{\dagger}_m - \frac{1}{2} \{J^{\dagger}_mJ_m, \rho\}\right) - \frac{1}{2} \sum^N_n\left\{ K_n, \rho\right\}
    \end{aligned}
    \label{Liou}
\end{equation}
Here, square (curly) brackets denote the (anti-) commutator. The first term accounts for the coherent evolution under Hamiltonian $H$ (in angular frequency units). The second term, $\hat{\hat{D}}\rho$, is in so-called Lindblad-form and describes decoherence processes in the Born-Markov approximation \cite{breuer2002theory}. The sum extends over $M$ quantum jump or collapse operators, $J_m$, the maximal number of which is one smaller than the square of the Hilbert space dimension \cite{huang2008non}. The third term, $\hat{\hat{K}}\rho$, here assumed in Haberkorn form, is unique to the treatment of the spin dynamics of radical systems. It describes chemical transformations of the radicals, i.e. their spin-selective recombination to form various reaction products. Typically, the $K_n$ relate to the singlet or triplet projection operators of the reactive pair ($i$,$j$) of radicals: 
\begin{equation}
\label{Kn}
    K_n = k^{(i, j)}_{S, T}P^{(i, j)}_{S, T},
\end{equation}
with
\begin{equation}
    P^{(i, j)}_S = \frac{1}{4} - \bf{S}_i \cdot \bf{S}_j, 
\end{equation}
and
\begin{equation}
    P^{(i, j)}_T = 1 - P^{(i, j)}_S,
\end{equation}
 where $S_i$ denotes the spin vector operator (in multiples of $\hbar$) of spin $i$. For recombination terms $K_n$ of the form of eq.\ \ref{Kn}, 
$K_n = K^{\dagger}_nK_n$ applies and the recombination operator assumes a form that, except for one term missing, is reminiscent of the Lindbladian. However, as a consequence of the presence of $\hat{\hat{K}}\rho$ the dynamics do not preserve the trace of the $\rho$, i.e.\ $\rho$ is actually the density operator weighted by the probability that the radical system has not yet recombined. In principle, this peculiarity can be avoided by introducing shelfing states, which allows one to recover the classical Lindblad form throughout \cite{gauger2011sustained}. However, this comes at the cost of enlarging the Hilbert space by product states, the spin degrees of freedom of which are assumed unobserved. Also note that by absorbing the singlet-triplet dephasing in the $J_m$s, eq.\ \ref{Liou} applies to various approaches to treat reactive radical systems, including the Jones-Hore model and the quantum master equation approach \cite{fay2018spin, jones2010spin, jones2011reaction}.
The equation of motion can be re-expressed in the form:
\begin{equation}
\label{eq21}
    \frac{d \rho}{dt} = -i\left(H_{\mathrm{eff}} \rho - \rho H^{\dagger}_{\mathrm{eff}} \right) + \sum^M_m J_m \rho J^{\dagger}_m,
\end{equation}
with $H_{\mathrm{eff}}$ denoting the effective Hamiltonian:
\begin{equation}
\label{Heff}
    H_{\mathrm{eff}} = H - \frac{i}{2}\sum^N_n K_n -\frac{i}{2} \sum^M_m J^{\dagger}_mJ_m,
\end{equation}
which marks the starting point of the MCWF approach. As $H_{\mathrm{eff}}$  is non-Hermitioan, it induces a non-structure-preserving map in the Hilbert space of the system. The algorithm starts out from an ensemble of state-vectors, $ \left\{\left| \phi(0) \right>\right\}$ , that appropriately sample the initial density operator $\rho(0)$. The state vectors are assumed to evolve under the effective Hamiltonian $H_{\mathrm{eff}}$ according to 
\begin{equation}
   \frac{d}{dt}  \left| \phi(t) \right> = -i H_{\mathrm{eff}}  \left| \phi(t) \right>
\end{equation}
and undergo occasional quantum jumps (to be described below). Closely following the exposition of \cite{molmer1993monte}, we shall consider the evolution of $ \left| \phi(t) \right>$ to $ \left| \phi(t + \delta t) \right>$, where the time increment $\delta t$ is arbitrary, but sufficiently small such that terms including  $\delta t$ of order 2 and higher can be neglected. Under this assumption, the non-unitary evolution produces the state:
\begin{equation}
\label{phi1}
     \left| \phi^{(1)}(t + \delta t) \right> \approx \left(1 - iH_{\mathrm{eff}}\delta t\right) \left| \phi(t) \right>.
\end{equation}
The square of the $l^2$ norm of the state decreases during this evolution from its value at $t$, $ \left< \phi(t) \right|\left. \phi(t) \right> = l^2$, to:
\begin{equation}
    \left< \phi^{(1)}(t + \delta t) \right|\left. \phi^{(1)}(t + \delta t) \right> = \left(1 - \delta p\right)l^2,
\end{equation}
where to first order in $\delta t$:
\begin{equation}
    \begin{aligned}
    \delta p &= \frac{i \delta t}{l^2}  \left< \phi(t) \right| H_{\mathrm{eff}} - H^{\dagger}_{\mathrm{eff}}  \left| \phi(t) \right>
    &\\
    &\equiv \sum^M_m \delta p_m + \sum^N_n \delta p'_n.
    \end{aligned}
\end{equation}
We later show that the squared norm $l^2$ cancels, so this is no significant divergence from the original approach, which assumes $l^2 = 1$. Yet, the more general assumption of $l^2 \neq 1$ was impelled here by the intrinsic non-trace preserving formulation of the dynamics in the presence of spin-selective recombination processes as described by eq.\ \ref{Liou}. $\delta p$ is interpreted as the probability that a quantum jump occurs within time interval $\delta t$. The probability that this jump involves the $m^{\text{th}}$ Lindblad term and the $n^{\text{th}}$ reaction term is given by $\delta p_m$ and $\delta p'_n$, respectively:
\begin{equation}
    \begin{aligned}
    \delta p_m &= \frac{\delta t}{l^2} \left< \phi(t) \right| J^{\dagger}_mJ_m \left| \phi(t) \right> \\
    &\\
    \delta p'_n &= \frac{\delta t}{2 l^2} \left< \phi(t) \right| K^{\dagger}_n + K_n \left| \phi(t) \right>\\
    &\\
    & = \frac{\delta t}{l^2} \left< \phi(t) \right| K_n \left| \phi(t) \right>,
    \end{aligned}
\end{equation}
where the last equality in the expression of $\delta p'_n$ applies for Hermitian $K_n$ (e.g.\ reaction terms composed from singlet and triplet projection operators). It is implied here that $\delta p << 1$ (as $\delta t$ is small), which guarantees that the probability of two jumps occurring within one timestep $\delta t$ is negligible. Quantum jumps are introduced into the time evolution of the state vectors as a stochastic element, thereby mimicking the physically expected, uncertainty of quantum processes. To this end, a quasi-random number $u$ between zero and one is drawn from the continuous uniform distribution. If, as in the vast majority of cases by construction, $\delta p < u$, no jump has occurred and the state vector is renormalized and propagated on. If, however, $\delta p \geq u$, a quantum jump is executed. The actual jump process is again selected at random from the $M$ Lindblad and $N$ kinetic terms, whereby the relative probability is 
$\delta p_m/\delta p$ and $\delta p'_n/\delta p$, respectively.
This leaves us with three distinct events to consider in the time-evolution of the state vector: no jump, Lindbladian jump, and recombination/reaction. In the event of no jump, which occurs with probability $1 - \delta p$ , the re-normalized state vector at 
$t + \delta t$, $ \left| \phi(t + \delta t) \right>$, can be chosen as
\begin{equation}
     \left| \phi(t + \delta t) \right>\Big|_\text{no jump} = l \frac{\left| \phi^{(1)}(t + \delta t) \right>}{||\left| \phi^{(1)}(t + \delta t) \right>||}.
\end{equation}
Alternatively, in the event of a jump associated with $J_m$ (occurring with probability $\delta p_m$), the re-normalized wavefunction is obtained from
\begin{equation}
     \left| \phi(t + \delta t) \right>\Big|_\text{jump} = l\frac{J_m \left|\phi(t) \right>}{||J_m \left|\phi(t) \right>||}. 
\end{equation}

So far, this entirely equals the established MCWF approaches. The new case that we have introduced is that of reaction/termination, which occurs with probability $\delta p'_n$. As the associated superoperator is not of Lindblad form, this event requires an approach that differs from MCWF algorithm as traditionally implemented. We physically reason that a termination reaction should eliminate the trajectory upon which it occurs, rather than propagating it. Thus we choose, for the termination state, the new state vector to be the zero function:
\begin{equation}
\label{zero}
    \left| \phi(t + \delta t) \right>\Big|_\text{reaction} = 0
\end{equation}
for $t + \delta t$ and all subsequent times. This is equivalent to stating that all expectation values following the reaction event are equated to zero, i.e.\ the quantum state is whence absent from the population weighted ensemble.

We demonstrate equivalence between this extended MCWF approach and the master equation approach by showing that the master equation, eq.\ \ref{Liou}, can be recovered from the ensemble average of MCWF trajectories. We begin by considering a pure state with density operator 
$\sigma(t) = \left| \phi(t) \right>\left< \phi(t) \right|$. We find this quantity at the later time $t + \delta t$ by averaging over many MCWF trajectories. This will give rise to the time averaged $\sigma(t + \delta t)$, here denoted $\bar{\sigma}(t + \delta t)$ , as a linear combination of the state-vector diads from above, each weighted by the associated probability:
\begin{equation}
\label{sigma}
\begin{aligned}
    \Bar{\sigma}(t + \delta t) &=  l^2 (1 - \delta p') \frac{\left|\phi^{(1)}(t + \delta t) \right> \left<\phi^{(1)}(t + \delta t) \right|}{||\left|\phi^{(1)}(t + \delta t) \right>||^2}\\
    &\\
    & + \sum_m l^2 \delta p_m\frac{J_m\left|\phi(t) \right> \left<\phi(t) \right|J^{\dagger}_m}{||J_m\left|\phi(t) \right>||^2}\\
    &\\
    & + \sum^N_n\delta p'_n \: 0.
\end{aligned}
\end{equation}
We have deliberately included $\delta p'_n$ multiplied by its zero generator to make clear the distinction between this method and the previous MCWF implementation. Using eqs.\ \ref{phi1} and \ref{zero}, this simplifies to:
\begin{equation}
\begin{aligned}
\Bar{\sigma}(t + \delta t) &= \left| \phi^{(1)}(t + \delta t) \right>\left< \phi^{(1)}(t + \delta t) \right| 
&\\
&+ \delta t \sum^M_m J_m\left|\phi(t) \right> \left<\phi(t) \right|J^{\dagger}_m,
\end{aligned}
\end{equation}
which can be written to first order in $\delta t$ as:
\begin{equation}
    \begin{aligned}
     &\Bar{\sigma}(t + \delta t) = \\
     &\\
     &\left( 1 - i H_{\mathrm{eff}} \delta t\right)\sigma(t)\left(1 + i H^{\dagger}_{\mathrm{eff}} \delta t \right)
     + \delta t\sum^M_m J_m \sigma(t) J^{\dagger}_m \\
     &\\
     &=\sigma(t) - i\delta t\left( H_{\mathrm{eff}}\bar{\sigma}(t) - \bar{\sigma}(t) H^{\dagger}_{\mathrm{eff}} \right) + \delta t\sum^M_m J_m \sigma(t) J^{\dagger}_m.
\end{aligned}  
\end{equation}
Finally, in the limit $\delta t \to 0$ we recover eq.\ \ref{eq21} for the pure initial state. The approach also holds for any convex combination of initial states, if the initial state is sampled from these states and is thus general. This completes the proof.

In summary, by imposing that a reaction event terminates the trajectory that is being propagated we have demonstrated the equivalence between an adapted version of the MCWF approach and the master equation that governs radical pair dynamics, or more generally, radical system dynamics, with recombination terms of non-Lindbladian form.

\section{Implementation}

Actually evaluating the MC evolution as a succession of many tiny steps $\delta t$, each of which treated to first-order in time as done above, is tedious. Instead, we uses the following algorithm to simulate a single realization of a quantum system \cite{plenio1998quantum}. Observe that based on the description from above, the probability of no-jump after $n$ steps is
\begin{equation}
    \begin{aligned}
    P_{no}(t &= n \delta t) = \Pi^{n-1}_{j=0} \norm{(1 - i H_{\mathrm{eff}} \delta t)\left|\phi(j \delta t)\right>}^2\\
    &\\
    & = \norm{\left(1 - i H_{\mathrm{eff}} \delta t \right)^n \left|\phi(0)\right>}^2.
    \end{aligned}
\end{equation}
In the limit $\delta t \to 0$, while $n \delta t = \tau$, this expression reduces to:
\begin{equation}
    P_{no}(\tau) = \norm{\exp(-i H_{\mathrm{eff}} \tau) \left|\phi(0)\right>}^2,
    \label{Pno}
\end{equation}
and the jump probability is:
\begin{equation}
    P(\tau) = 1 - P_{no}(\tau) = 1 - \norm{\exp(-i H_{\mathrm{eff}} \tau) \left|\phi(0)\right>}^2.
\end{equation}
This can be interpreted as the cumulative distribution function of the waiting time $\tau$ until the next jump. Thus, instead of painstakingly accumulate many $\delta t$-steps, we sample from the waiting time distribution. Practically, this can be implemented by drawing a uniform random number $u \in [0, 1)$ and performing the non-unitary deterministic time evolution generated by $H_{\mathrm{eff}}$ until the squared norm of the state vector falls below $u$. At the moment where this happens, a quantum jump is executed. The collapse/kinetic operator for the jump is selected from the $N + M$ possibilities at random, where the relative probability of choosing $J_m$ and $K_n$ is proportional to $\delta p_m$ and $\delta p'_n$, respectively. This approach increases the efficiency of solving the problem as it allows one to rely on higher order ordinary differential equation (ODE) solvers (rather than the inefficient Euler method used in the derivation above), which utilize markedly larger and, possibly, adaptive time steps. As the approach is derived from the $\delta t \to 0$ limit, it also avoids the problem of quantum jumps taking finite time (see discussion in \cite{kornyik2019monte}). The downside is that exact time point of the quantum jump has to be identified. Using state-of-the-art ODE solvers that provide dense outputs, i.e. representations of the solution as interpolation function in $t$, this can be realized efficiently by backtracking the solution to the point where the jump condition is met \cite{rackauckas2017differentialequations}. We find that the problems considered here are efficiently integrated by $5/4$ Runge-Kutta methods, such as Dormand-Prince's and Tsitouras’ approach \cite{dormand1986reconsideration, tsitouras2011runge}, both of which provide free 4$^\mathrm{th}$ order interpolants to be used in backtracking and interpolating the solution on a user-provided time grid (we have used a point every nanosecond). The performance of the Runge-Kutta methods could even be exceeded by using adaptive multi-step methods. In particular, the 5$^\mathrm{th}$ order Adams-Moulton method (using the 4$^\mathrm{th}$ order Runge-Kutta approach to calculate starting values) proved highly efficient \cite{schwarz2013numerische}. 

We consider radical pairs that are generated in the singlet electronic state. The initial state is
thus an incoherent mixture of singlet states whereby all nuclear spin configurations are equally
likely, as the states are quasi-degenerate and the temperature is comparably high:
\begin{equation}
    \rho(0) = \frac{1}{Z} P_s \otimes_i 1_i
\end{equation}
Here, $Z$ is the total number of nuclear spin states and the direct product comprises all nuclei via
index $n$. Consequently, the MCWF approach requires sampling different nuclear spin states in a way
that converges, for every nuclear Hilbert space, to the required scaled unity density $1_i$. This
could obviously be realized by iterating over all $Z$ nuclear spin states e.g.\ in the canonical basis (i.e.\ the eigenstates of $I_i^2$ and $I_{i,z}$). However, the exponential scaling of $Z$ with the number of nuclear spin renders this approach impractical for large spin systems, i.e.\ those that the MCWF approach
aims to cover. This exponential scaling can be avoided by, once again, performing a Monte Carlo
sampling over randomly chosen initial states \cite{weisse2006kernel}. A study by the Manolopoulos group suggests that, in the context of the wavefunction-based
approach to solving eq.\ \ref{Liou} without Lindblad dissipators, this limitation can be overcome by
stochastically sampling the nuclear spin wave functions in the form of spin coherent states \cite{lewis2016efficient}. This approach promises rapid convergence of expectation values with the number of sampled states and, importantly, the number of required states not scaling (or even decreasing) with the size of the Hilbert space \cite{weisse2006kernel}. The over-complete set of spin coherent state, parametrized by $\theta$ and $\phi$, can formally be obtained by rotating the $I_{i,z}$ eigenstate with highest projection, $\left|I_i, I_i\right>$, to point in the direction
$\hat{n} = \hat{n}(\theta, \phi)$ of polar angle $\theta$ and azimuthal angle $\phi$, e.g. by rotating by angle $\theta$ around the axis parallel to $\hat{n} \times \hat{z}$ (the hat on classical vectors is meant to mean normalisation) \cite{lieb1973classical, radcliffe1971some}:
\begin{equation}
\begin{aligned}
    &\left| \Omega_i\right> =  \left|\theta_i, \phi_i\right> \\
    & = \exp\left( i \theta \widehat{\hat n \times \hat z} \cdot \vec{I_i}  \right) \left|I_i, I_i\right>\\
    & = cos\left(\frac{\theta_i}{2}\right)^{2I_i} \exp\left(tan\left(\frac{\theta_i}{2}\right)e^{i\phi}I_{i, -}\right) \left|I_i, I_i\right>.
\end{aligned}
\end{equation}
As the states satisfy the completeness relation
\begin{equation}
    1_i = \int^\pi_0 sin(\theta_i)d\theta_i\int^{2\pi}_0 d\phi_i \left|\Omega_i\right>\left<\Omega_i\right|,
\end{equation}
the MCWF approach can be realized by randomly sampling the orientation $\hat n$ for each of the nuclear spins from the unit sphere. As the MCWF is stochastic as such, the additional spin coherent states sampling is seamlessly integrated.

\section{Results}

\begin{figure}[tb]
    \centering
    \includegraphics[width=7cm]{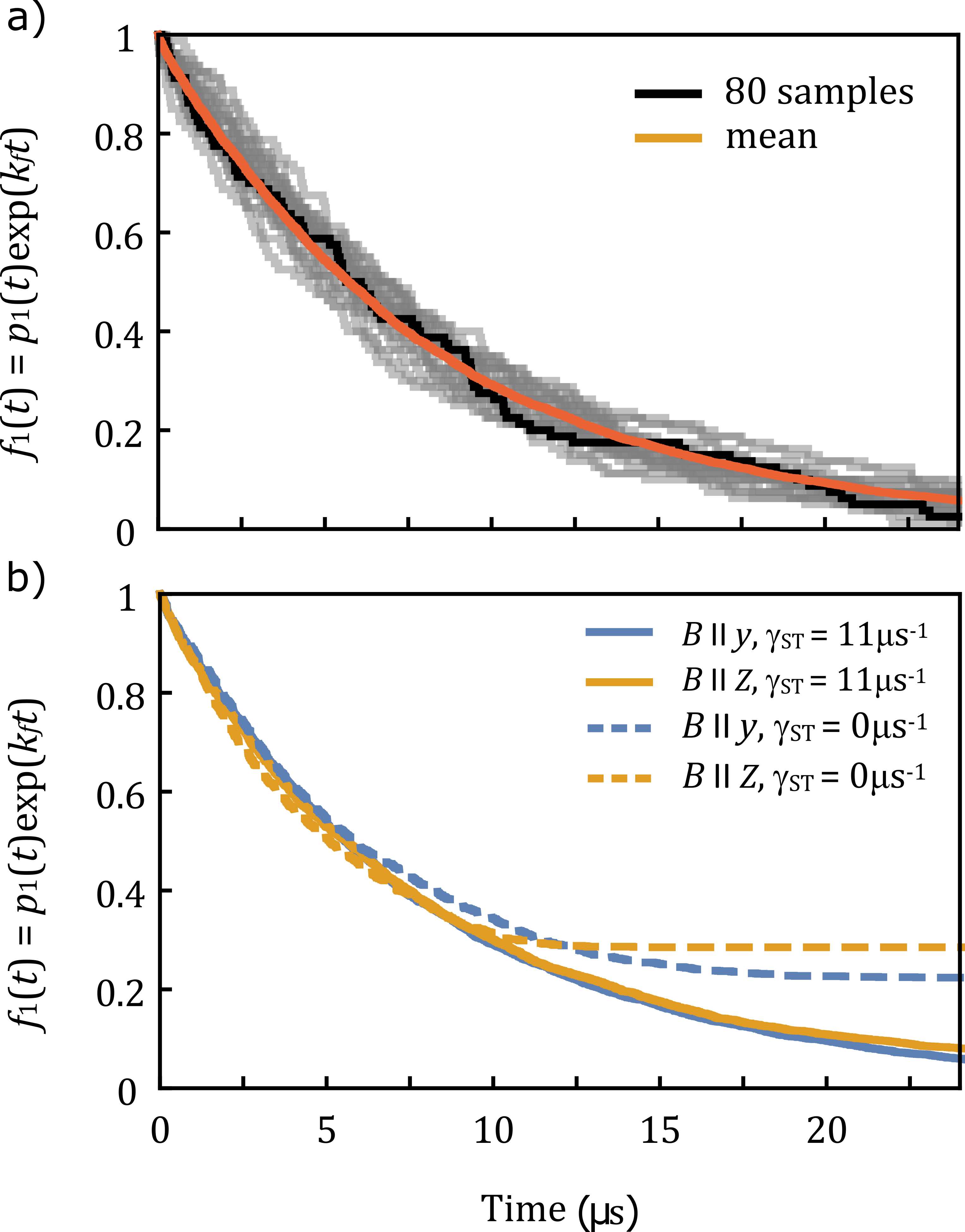}
    \caption{Transformed survival probability, $f_1(1) = p_1(t) \exp(k_f t)$, as a function of time for the [FAD$^{\bullet-}$ W$^{\bullet+}$] system with 8 + 8 nuclear spins, subject to S/T-dephasing at the rate $\gamma_{ST} = 11$ \si{\us}$^{-1}$ and a magnetic field of $B = 1~$mT. A recombination rate of $k_b = 0.5$ \si{\us}$^{-1}$ was assumed. (a) shows the reconstruction of the time-dependence of $f_1(t)$ for many trajectories using the MCWF approach. The solid red line shows the eventually converged mean; every gray curve, a single of which has been highlighted in black, corresponds to 80 samples. Here, $B = 1$ mT is aligned with the $\hat{z}$-axis of FAD. (b) summarizes the converged results for or different field directions in the absence and presence of S/T-dephasing. Note that S/T-dephasing strongly attenuates the anisotropy of the response to magnetic field. }
    \label{1}
\end{figure}

We have implemented the MCWF approach for spin dynamic calculations on radical pairs as outlined above. Here, we present an assessment of its performance established in terms of two prototypical radical pair systems with putative relevance to magnetoreception \cite{lee2014alternative, hiscock2016quantum, maeda2012magnetically}. These radical pairs involve a semi-reduced flavin adenine dinucleotide, FAD, non-covalently bound in the protein cryptochrome, i.e.\ FAD$^{\bullet-}$, and a partner radical. \textit{In vitro}, the combination of FAD$^{\bullet-}$ and an oxidized tryptophan radical, W$^{\bullet+}$, is known to convey magnetosensitivity \cite{maeda2012magnetically, kattnig2016chemical}. \textit{In vivo}, the identity of the second radical is less clear and currently fiercely debated \cite{kattnig2017radical}. In model calculations, systems of so called reference-probe character have been found to elicit large anisotropic magnetic field effects \cite{lee2014alternative, procopio2020reference}. The prototypical system of this kind is [FAD$^{\bullet-}$ Z$^{\bullet}$], where the flavin is combined with a radical devoid of hyperfine interactions, Z$^{\bullet}$. A radical of this kind could possibly result from a reoxidation reaction of the fully reduced FAD cofactor with molecular oxygen \cite{hammad2020cryptochrome, pooam2019magnetic, wiltschko2016light}. Note that many details of cryptochrome magnetoreception are as yet unknown and no definite picture has emerged from the combined literature. However, this shall not burden us here, where the purpose is to discuss MCWF approach for recombining radical pairs as treated within the Haberkorn, respectively quantum master equation, framework. Radical pair magnetic field effects are attributed to a delicate interplay of the coherent evolution, predominantly due to hyperfine and the Zeeman interaction, and spin selective recombination. For oriented systems, the typical Hamiltonian is of the form
\begin{equation}
    H = H_1 + H_2,
\end{equation}
where
\begin{equation}
    H_k = \frac{\mu_B g_j}{\hbar} \vec{B} \cdot \vec{S}_k +  \sum_{i} \vec{S}_k \cdot A_{ki} \cdot \vec{I}_{k,i}. 
\label{H}    
\end{equation}
Here, $\vec{B}$ is the applied magnetic field, $g_k$ the g-factor ($g$-anisotropy is neglected here, because we focus on organic radicals in comparably weak magnetic fields), $A_{ki}$ is the hyperfine tensor and $\vec{S}_k$ and $\vec{I}_{k,i}$ are the individual electron and nuclear spin operators, respectively. In addition, the exchange and electron-electron dipolar interaction should be considered. These interactions are sometimes the source of unexpected effects, but often are found to attenuate MFEs in weak magnetic fields \cite{babcock2020electron, efimova2008role}. As is commonly done, we shall neglect these extra terms here for simplicity. Note nonetheless that their inclusion would not pose an additional difficulty as the effective Hamiltonian, $H_{\mathrm{eff}}$, as resulting from eq.\ \ref{H} does not in general allow for a factorisation anyway. The radical pair systems considered here can either recombine in the singlet configuration (with rate $k_b$) or proceed in a non-spin selective process to a reaction product (rate $k_f$), which in the context of magnetoreception is thought to involve a protein structure rearrangement, whence innervating a signalling cascade \cite{kattnig2018molecular}. The kinetic superoperator $\hat{\hat{K}}$ is thus of the form
\begin{equation}
    \hat{\hat{K}}\rho = -\frac{k_b}{2}\left\{P_S, \rho \right\} - k_f \rho,
\label{kf}    
\end{equation}
which is tantamount to setting $k_S = k_f + k_b$ and $k_T = k_f$ in eq.\ \ref{K}. We assume that the radical pairs are subject to S/T-dephasing (see eq.\ \ref{Kprime} above; dissipation rate $k_{ST} = \gamma_{ST}$) as a result of the reactive encounter process and/or the modulation of the exchange and electron-electron dipolar interaction terms by molecular motion. In addition, we assume random field relaxation, which is accounted for by a Lindblad dissipator with uncorrelated noise associated with the Cartesian spin operators of the two radicals $J_m \in \{S_{k,x}, S_{k,y}, S_{k,z}\}$ \cite{kattnig2016electron2, kattnig2017radical}. Assuming that the dissipation rates for all directions are equal to $\gamma_{RF,k}$, this gives rise to the combined term for radical $k$ of the form
\begin{equation}
    \hat{\hat{R}}_k\rho = \gamma_{RF, k} \left[ \sum_{\alpha \in \{x,y,z\}} S_{k, \alpha}\rho S_{k, \alpha} -\frac{3}{4}\rho  \right].
\label{R}    
\end{equation}

Furthermore, we assume that the radical pair is generated in the singlet state, e.g.\ by a swift, spin-conserving electron transfer reaction of diamagnetic precursors, and, thus, $\rho(0) = P_S/Tr[P_S]$.

We follow the singlet probability, $p_S(t) = Tr[P_S \rho(t)]$ or the survival probability, $p_1(t) = Tr[\rho(t)]$ of the radical pair over time, from which the yields of the recombination, $Y_S$ and escape/signalling product, $Y_1$ can be calculated from
\begin{equation}
    Y_S = k_b\int^\infty_0 p_s(t) dt,
\end{equation}
and
\begin{equation}
    Y_1 = k_f\int^\infty_0 p_1(t) dt,
\end{equation}
respectively. Evaluating $p_1(t)$ from the MCWF approach is particularly straight forward, as it is derived from sampling the recombination time only. In particular, no evaluation of expectation values on a regular time grid (in addition to the usual following of the trace of $\rho$ for the purpose of identifying the moments of quantum jumps) is necessary. This allows for a particularly efficient implementation.

\begin{figure}[htp]
    \centering
    \includegraphics[width=7cm]{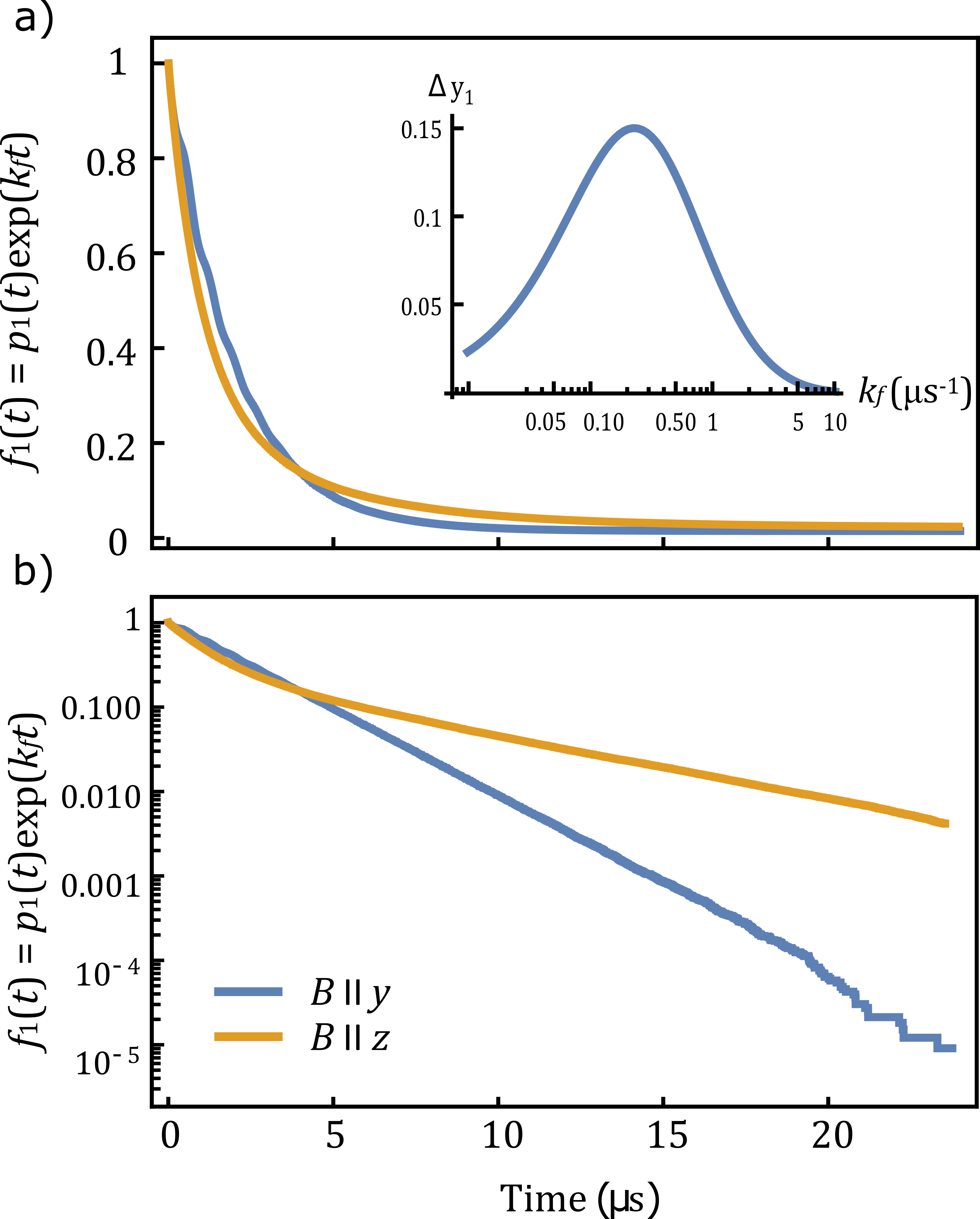}
    \caption{(a) 
    The test case used here was the [FAD$^{\bullet-}$ Z$^{\bullet}$] system, with $14$ spins under random field relaxation with rate $\gamma_{RF} = 0.2$ \si{\us}$^{-1}$ and $k_b = 2$ \si{\us}$^{-1}$. Here, we have assumed $B = 50$ \si{\micro\tesla}, which is of the order of the geomagnetic field at mid-latitude. The insert shows the dependence of the magnetic anisotropy for this system, evaluated as the difference of $Y_1$ when the field is in the $\hat{y}$ and $\hat{z}$-direction, respectively, on the forward rate constant $k_f$. (b) shows $f_1(t)$ on a logarithmic scale. $t_{\mathrm{max}} = 24$ \si{\us} and $N = 476,800$ Monte Carlo samples have been collected.}
    \label{2}
\end{figure}

Here, we report on our attempts of applying the MCWF approach to the two radical pair systems [FAD$^{\bullet-}$ W$^{\bullet+}$] and [FAD$^{\bullet-}$ Z$^{\bullet}$] subject to the outlined scenario, whereby a variable number of hyperfine-coupled nuclei was taken into account. The relevant hyperfine parameters, up to $12$ for FAD$^{\bullet-}$ (including two nitrogens with $I = 1$) and up to $8$ (one nitrogen) for W$^{\bullet+}$, are summarized in the Supporting Information to this text. For the simulations with a variable number of hyperfine-coupled nuclear spins reported here, the hyperfine interactions in FAD$^{\bullet-}$ were added in the order N5, N10, H6, 3 x H8, H$\beta$1, H$\beta$2, H9, and 3 x H7. For W$^{\bullet+}$ the order was N1, H1, H$\beta$2, H4, H2, H6, H$\beta$1, H7 and H5. Figure \ref{1} shows exemplary results for [FAD$^{\bullet-}$ W$^{\bullet+}$] subject to S/T-dephasing at the rate $\gamma_{ST} = 11$ \si{\us}$^{-1}$ and a magnetic field of $B = 1$ mT. Data for $B = 0$ mT are shown in the Supporting Information (Fig.\ S1). A recombination rate of $k_b = 0.5$ \si{\us}$^{-1}$ was assumed, which is typical for this kind of system. We show the transformed survival probability $f_1(t) = p_1(t) \exp(+k_f t)$ as a function of time, which is independent of $k_f$ as by the form of $\hat{\hat{K}}$, eq.\ \ref{kf}, the forward reaction induces a simple exponential decay of the density matrix. Typical values of $k_f$ would be of the order of $1$ \si{\us}$^{-1}$. Figure \ref{1}(a) shows how the MCWF approach re-constructs the time-dependence of an observable from many trajectories; 80 individual runs are combined in one grey line; the eventually converged average is shown as red solid line. Figure \ref{1}(b) shows converged results for different orientations of the magnetic field with and without S/T-dephasing. Note that in the absence of S/T-dephasing the dynamics populate states which are part of the kernel of the Liouvillian and thus do not decay. S/T-dephasing breaks the longevity of the associated radical pair population. Figure \ref{2} features the [FAD$^{\bullet-}$ Z$^{\bullet}$] spin system with $14$ spins under random field relaxation (cf.\ eq.\ \ref{R}) with rate $\gamma_{RF} = 0.2$ \si{\us}$^{-1}$. Here, we have assumed $B = 50$ \si{\micro\tesla}, which is of the order of the geomagnetic field at mid-latitude. Figure \ref{2}(a) shows $f_1(t)$ for different orientations of the magnetic field and the dependence of the magnetic anisotropy, assessed in terms of the difference of the escape yield for the magnetic field aligned with the $\hat{z}$ and $\hat{y}$-direction, on the rate $k_f$. To this end, the yield was obtained by numerically integrating $p_1(t)$ up to a cut-off time of $t_{max} = 24$ \si{\us}. $f_1(t)$ for a smaller number of samples and the convergence behaviour of $\Delta Y_1$ for this system are shown in the Supporting Information (Figures S2 and S3).

\begin{figure}[t]
    \centering
    \includegraphics[width=7cm]{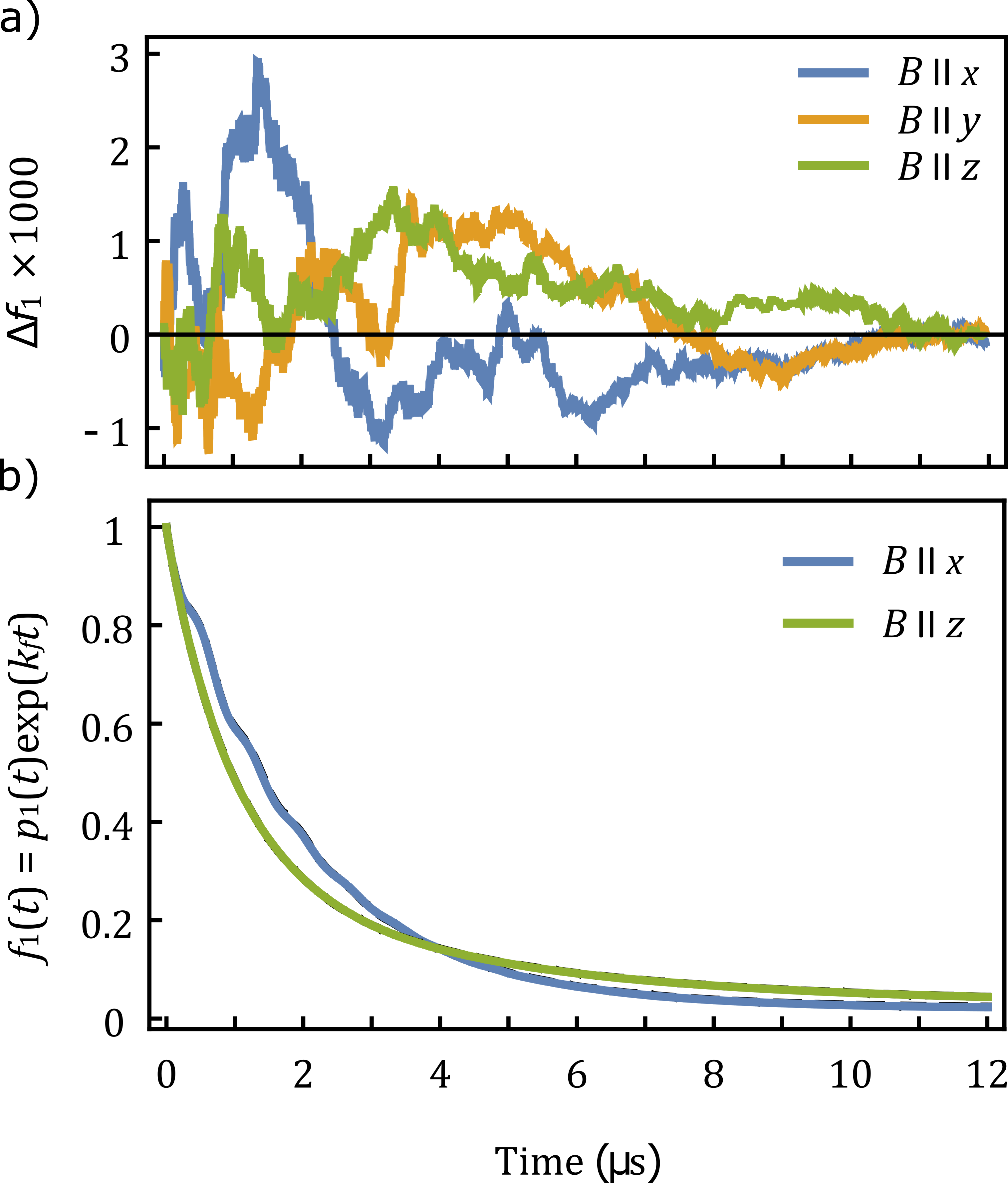}
    \caption{(a) Deviation of the direct ME integration and MCWF approach with $160,000$ samples drawn, for the [FAD$^{\bullet-}$ Z$^{\bullet}$] problem with random field relaxation ($\gamma_{RS} = 0.2$ \si{\us}$^{-1}$), $k_b = 2$ \si{\us}$^{-1}$ and $10$ coupled nuclei (including two nitrogen atoms) for different orientations of the magnetic field ($B = 50$ \si{\micro\tesla}) as indicated.
    (b) The transformed survival probability of this system as a function of time. The ME and MCWF approaches are indistinguishable on the image scale.}
    \label{3a}
\end{figure}

\begin{figure}[t]
    \centering
    \includegraphics[width=7cm]{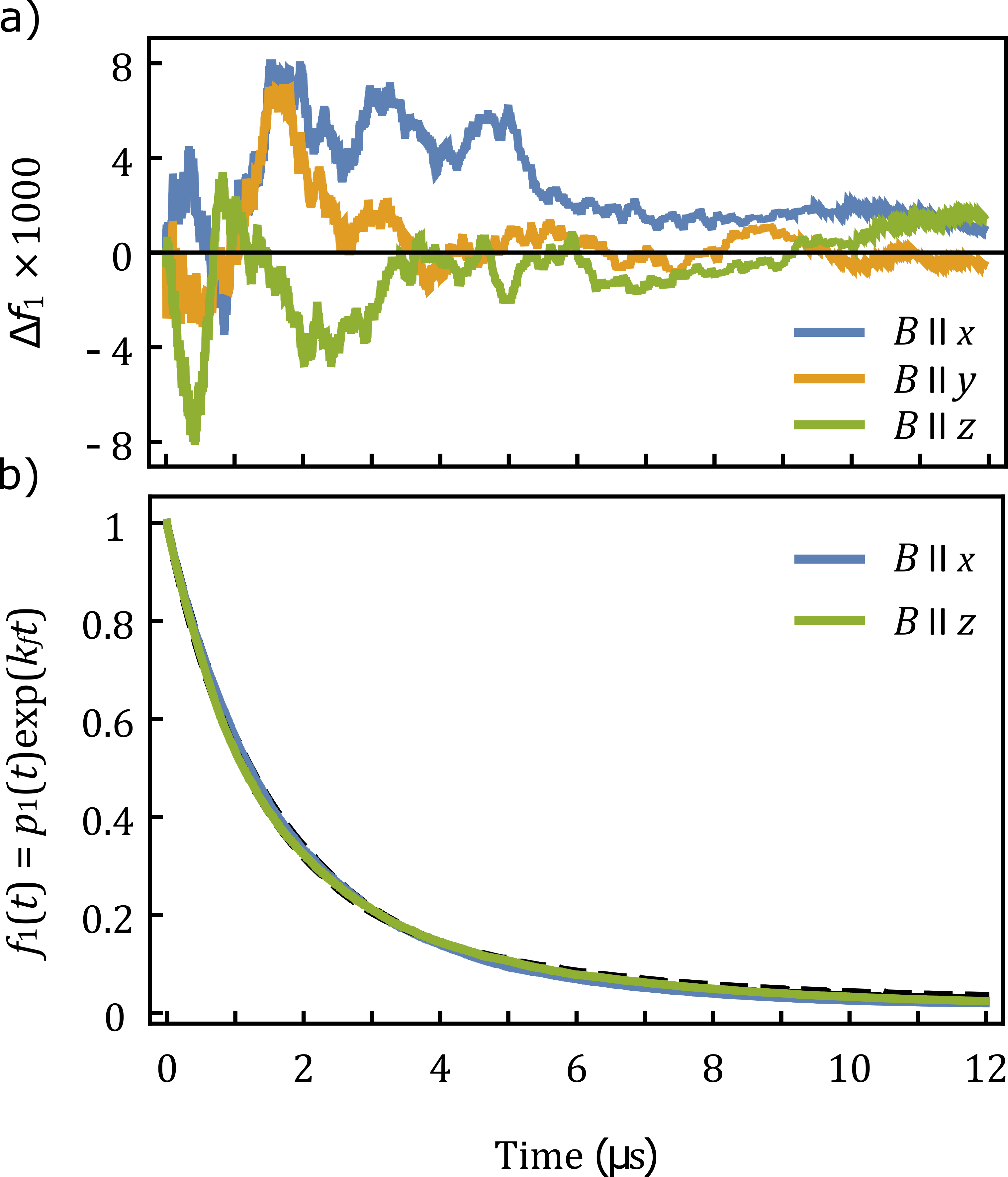}
    \caption{(a) Deviation of the direct ME integration and MCWF approach with $16,000$ samples drawn, for the [FAD$^{\bullet-}$ W$^{\bullet+}$] with $4$ hyperfine-coupled nuclear spin in every radical, random field relaxation ($\gamma_{RS} = 0.2$ \si{\us}$^{-1}$), $B = 1$ \si{\milli\tesla} and $k_b = 2$ \si{\us}$^{-1}$.
    (b) The transformed survival probability of this system as a function of time.}
    \label{3c}
\end{figure}

\begin{figure}[htp]
    \centering
    \includegraphics[width=8cm]{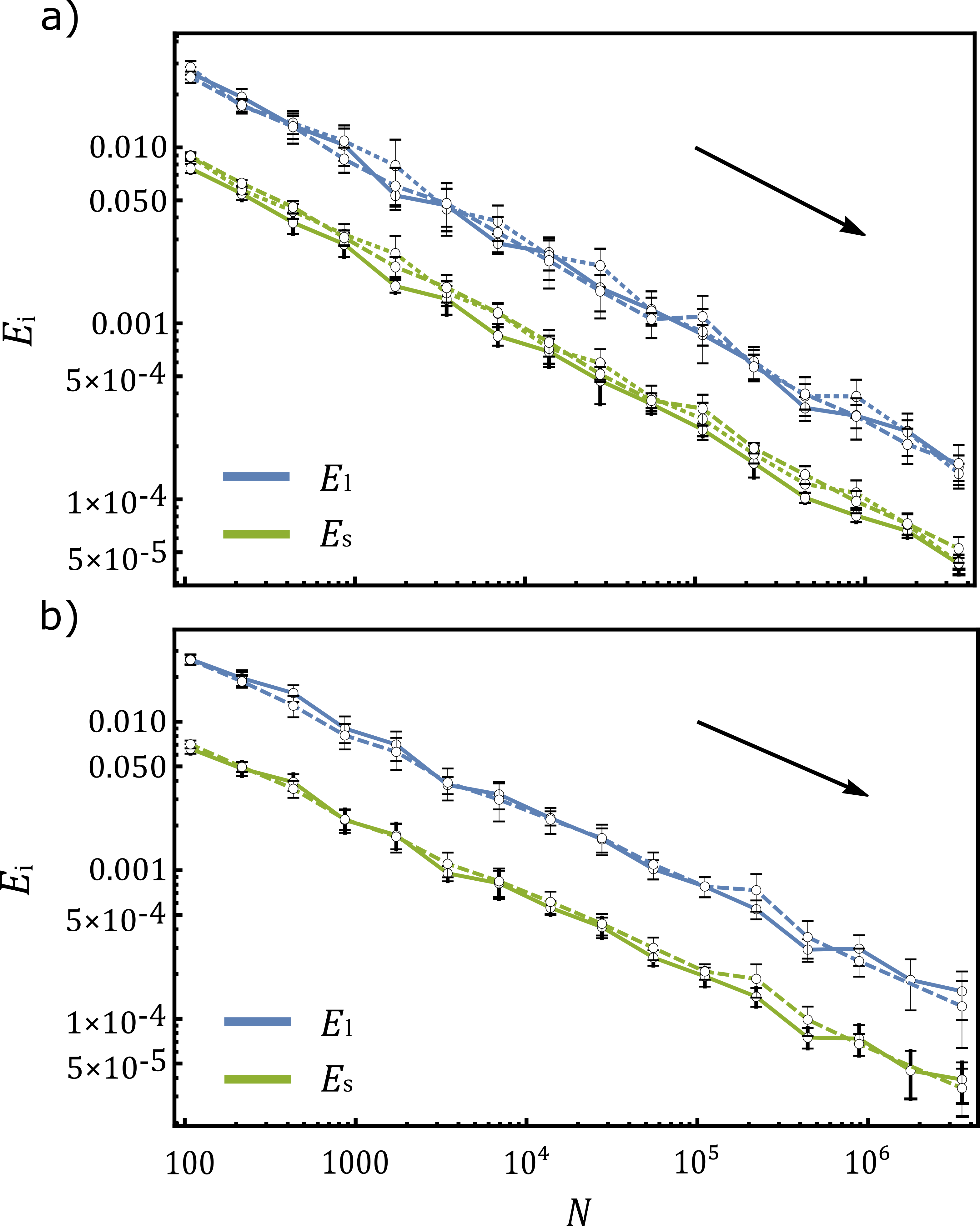}
    \caption{RMS error of the transformed singlet and  survival probability within the implementation of the MCWF approach compared to the numerical implementation of the direct integration of the master equation as a function of the number of Monte Carlo samples $N$. (a) applies to the 5 spin system FAD$^{\bullet-}$ W$^{\bullet+}$] system; (b) collects data for an 8 spin system of the same type. The errors $E_S$ and $E_1$ are shown in green and blue, respectively. Different sampling strategies of the initial nuclear spin configurations are encoded by line styles: solid lines: spin coherent state sampling; dashed lines: random sampling of nuclear spin states in the Zeeman basis; and dotted lines: complete sampling. The arrow indicates the slope of the expected $N^{-1/2}$-dependence. Linear fits to the data are in agreement with this expectation within statistical error. The error bars indicate two standard deviations of the mean error $E_i$ evaluated from 94 to 4 independent repeats for a given sample size $N$. All additional parameters are as for Fig.\ \ref{3a}.}
    \label{5}
\end{figure}

For smaller spin systems, we have compared the accuracy of the method to the result obtained from direct integration of the master equation using Tsitouras’ Runge–Kutta pairs of order $5(4)$ with adaptive time stepping to ascertain a relative and absolute error of $10^{-8}$. For the [FAD$^{\bullet-}$ Z$^{\bullet}$] problem with random field relaxation ($\gamma_{RS} = 0.2$ \si{\us}$^{-1}$) and $10$ coupled nuclei (including two nitrogen atoms), Figure \ref{3a} illustrates the deviation of the direct integration and the MCWF approach when $160,000$ Monte Carlo samples are drawn. The maximal deviations are of the order of $10^{-3}$ -- invisible to the eye when comparing $f_1(t) = p_1(t) \exp(k_b t)$ in the range from $0$ to $1$. Figure \ref{3c} shows similar data for [FAD$^{\bullet-}$ W$^{\bullet+}$] with $4$ hyperfine-coupled nuclear spin in every radical when $16,000$, ten times fewer, samples are averaged in the MCWF approach. Analogous data for S/T-dephasing are provided in the Supporting Information (Figure S4). These data show that the two approaches provide congruent results. Naturally, the accuracy of the MCWF method depends on the number of accumulated trajectories. Figure \ref{5} shows the root-mean-square error
\begin{equation}
    E_i = \sqrt{\frac{1}{t_{max}} \int^{t_{max}}_0 \left(f_{i, MCWF} - f_{i, ME}\right)^2 dt}
\label{E}    
\end{equation}

\noindent of $f_i(t) = p_i(t) \exp(k_f t)$ as a function of the number of MC samples. ME stands for the direct, i.e. naïve, integration of the master equation. $t_{max}$ was set to $24$ \si{\us}. The error bars indicate two times the standard deviation of the mean of $E_i$ evaluated from $4$ to $94$ independent repeats (depending on the sample number) of the error calculation. Both $f_1$ and $f_S$ decrease approximately with $N^{-1/2}$, as is expected for the standard error of Monte Carlo estimators. Importantly, for the studied systems, no systematic difference between the MCWF and ME method became apparent, which is expected for an implementation based on eq.\ \ref{Pno} provided that the error tolerances associated with the numerical integration and quantum jump time localisation are chosen sufficiently low. We have also compared different approaches of sampling the initial nuclear spin wavefunction. In addition to the spin coherent state sampling described above, we have randomly picked nuclear spin wave function in the $\{I_{i}^2,I_{i,z}\}$-basis and used a complete set of basis functions, i.e. all $\{I_{i}^2,I_{i,z}\}$-basis states were sampled in succession. We could not discern a significant difference of the error of $f_1(t)$, i.e.\ the standard deviation associated with the sampling process (which is larger than the indicated standard deviations of the mean of the sampled error) exceeded the differences of the approaches. This indifference is reassuring, as it suggests that the error introduced by stochastically sampling the nuclear spin functions does not lead to undue error compared with complete sampling, where it is possible. The approach based on spin coherent states is valuable as for large spin systems drawing less than $Z$ samples is often unavoidable or, in fact, desirable, which is where this method is expected to offer good convergence \cite{lewis2016efficient}. Based on the results here, however, we cannot answer the question of whether spin coherent state sampling is superior to the random picking of the finite set of states. For $f_S$ the analogous conclusions apply. It is interesting to note that for the system analysed here the ratio of the errors $E_S/E_1$ appears to be systematically smaller for the spin coherent state sampling than the other implemented approaches (whilst the differences are again smaller than the standard deviations). This might indicate a small inherent advantage of the spin coherent state when singlet yields are observed. This observation will require more detailed studies to substantiate.

\begin{figure}[htp]
    \centering
    \includegraphics[width=8cm]{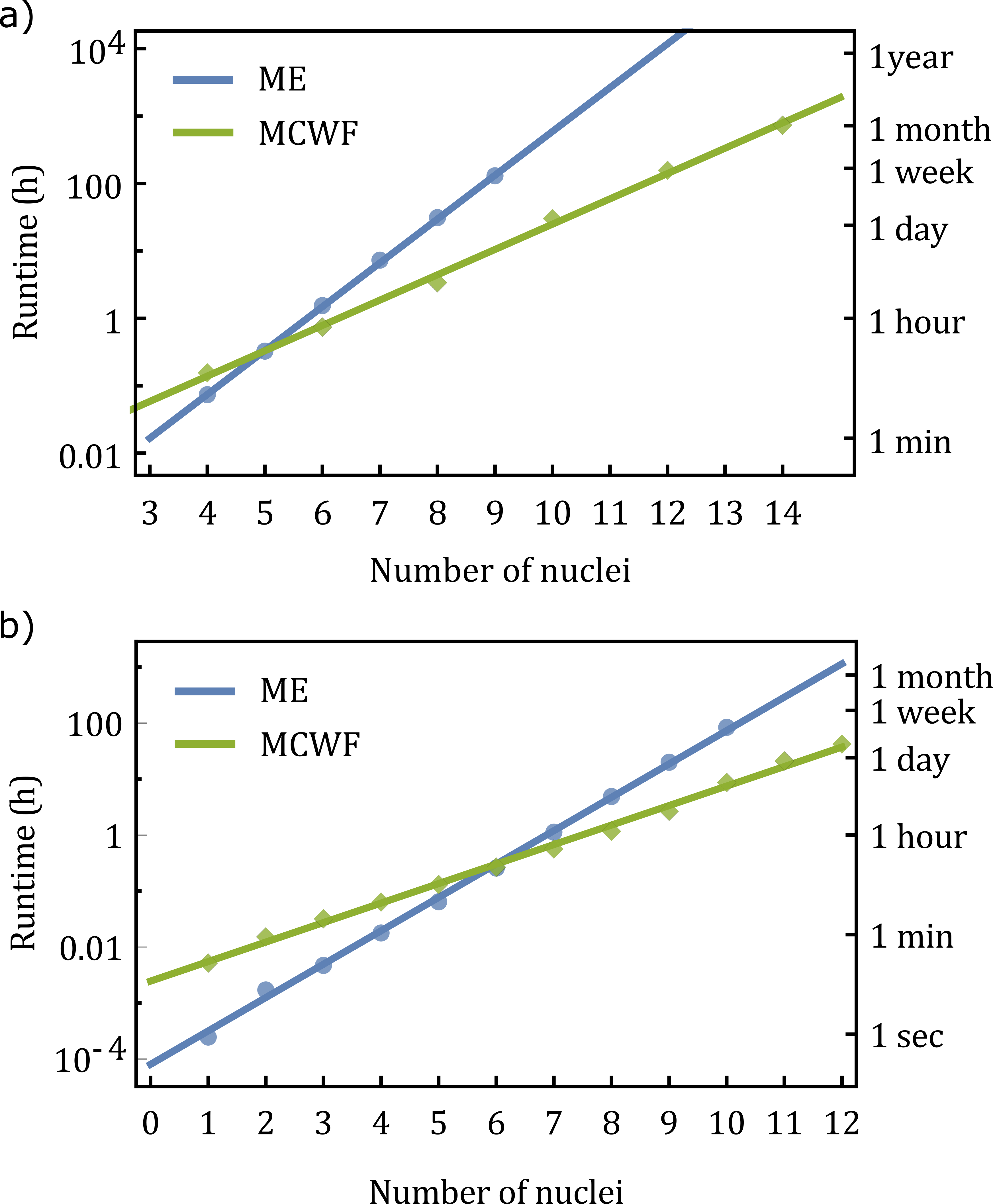}
    \caption{Comparison of the CPU time required to generate a solution for both the integration of the ME and the implementation of the MCWF approach, as a function of the number of coupled spins in the system. Panel (a) shows a [FAD$^{\bullet-}$ W$^{\bullet+}$] system that comprised all nitrogen spins and a variable number of proton spins, and panel (b) shows a basic [FAD$^{\bullet-}$ Z$^{\bullet}$] system. Simulation parameters are as given in Fig.\ 1 and 2 for the respective systems. $2^{16}$ samples have been drawn for the MCWF approach and $t_{max} = 12$ \si{\us}. The MCWF method used the Dormand-Prince 5/4 Runge-Kutta method with an absolute error tolerance of $10^{-8}$ and a relative error tolerance of $10^{-6}$.}
    \label{4}
\end{figure}

The MCWF approach strength lies in its applicability to comparably large spin system for which the ME approach cannot realistically be applied. This advantage becomes obvious when comparing the CPU times to generate a solution for the two approaches as a function of the number of coupled spins, as it is summarized in Figure \ref{4}, with a further result for the S/T-dephasing scenario shown in the Supporting Information (Figure S5). Here, we have started out from a basic [FAD$^{\bullet-}$ Z$^{\bullet}$] or [FAD$^{\bullet-}$ W$^{\bullet+}$] system that comprised all nitrogen spins. Adding one proton spin at a time, the runtime dependence as given in the figure is obtained. For the ME approach, the elapsed time scales according to $\mathcal{O}(4^n)$ to $\mathcal{O}(4.5^n)$, where $n$ is the number of considered nuclear spins, while for the MCWF approach scales as $\mathcal{O}(2.4^n)$. Overall, the $\mathcal{O}(4^n)$ scaling of the ME method quickly renders the calculation formidable. E.g.\ for only $4 + 4$ nuclear spins, the integration of the [FAD$^{\bullet-}$ W$^{\bullet+}$] system up to $12$ \si{\us} already requires $31$ hours. The direct integration of the systems shown in Fig.\ \ref{1} and \ref{2} can be considered intractable by the current means as it would require $48$ days for the smaller [FAD$^{\bullet-}$ Z$^{\bullet}$] system or even $453$ years for the [FAD$^{\bullet-}$ W$^{\bullet+}$] system (provided the memory requirements could be met). On the other hand, the weaker $\mathcal{O}(2.4^n)$ of the MCWF approach found here allows much larger spins systems, realistically up to ~$20$ spins, to be integrated. Here, the key question is not only the Hilbert space dimension but in addition the required accuracy of the solution. If high accuracies are required, the MCWF method can also be arbitrarily expensive as the $\mathcal{O}(N^{1/2})$ scaling of the error of the averaged quantities mandates potentially huge sample sizes. For this reason, for small systems, the direct integration is preferred.

\section{Discussion}

The great advantage of the Monte Carlo wavefunction method for obtaining time-dependent quantum expectation values is that systems of greater complexity are amenable to this treatment than is possible using the direct integration of the master equation. This advantage stems from the fact that the method rests on the propagation of wavefunctions. 
For a Hilbert space dimension $d = 4Z$, the number of wave function components is equal to $4Z$ while the number of density matrix components is equal to $d^2 = 16Z^2$. In the worst case, i.e.\ for dense operators, the propagation would scale quadratically in the number of components. Practically, the sparsity of $H_{\mathrm{eff}}$ caters for a more favourable scaling (for both approaches). In particular, we expect an ideal scaling of $\mathcal{O}(d ln d)$ for the propagating of a state vector, as a) the bilinear combination of spin operators of the form $S_{k,\alpha}I_{ki,\beta}$ (with $\{\alpha, \beta\} \in {x,y,z}$) directly couple at most five states and that b) there are $\mathcal{O}(ln d)$ such terms. Regardless of the actual scaling, the MCWF propagation will turn out to be exceedingly more efficient than directly integrating the master equation, as the latter requires $\mathcal{O}(d)$ applications of $H_{\mathrm{eff}}$ and therefore $\mathcal{O}(d^2 ln d)$ operations. Practically, we found a scaling of $\mathcal{O}(d^{1.26})$ for the MCWF method applied to the [FAD$^{\bullet-}$ W$^{\bullet+}$] radical pair and $\mathcal{O}(d^{2.2})$ for the direct integration. 

The primary disadvantage of the MCWF method is that the calculated quantities contain a statistical uncertainty, which needs to be reduced to an application-specific limit. The uncertainty results from the variability of the initial nuclear spin configuration and the stochasticity of the quantum jumps. It has to be contained by sampling, which however, comes at a significant cost of computation time, as the statistical error decreases as $N^{-1/2}$ with increasing number of samples $N$. Fortunately, the method inherits a remarkable scaling behaviour from wavefunction-based approaches to spin dynamics for closed quantum system, which suggest that for large spin systems significantly fewer than $Z$ samples of the initial nuclear spin configuration are often sufficient to arrive at adequately converged observable trajectories \cite{lewis2014asymmetric}. Practically, a constant number of samples, independent of problem dimension, proves to work well, suggesting that the scaling behaviour as suggested above still applies to the MCWF approach on the whole. For more than approximately 10 spins, the MCWF method quickly becomes the only feasible approach to integrate eq.\ \ref{Liou}. For small systems, on the other hand, the added overhead of averaging a large number of stochastic trajectories to obtain the open system dynamics, outweighs the benefit of a moderate memory saving. Master equation methods are therefore generally more efficient when Hilbert space dimensions are on the order of a couple of hundred states or smaller.

Alternative approaches to simulate the dynamics of relaxing systems could in principle be built on the closed-system dynamics of spin systems. To this end, one would have to engineer a time-dependent stochastic process such that the coherent dynamics sampled over this process will give rise to the dynamics as predicted by the master equation \ref{Liou} \cite{lewis2016efficient}. Up to second order, this could e.g.\ be realized in the framework of the Redfield approach \cite{breuer2002theory}. However, this strategy does not overcome the problem that many trajectories have to be sampled and comes at the conceptual disadvantage that a particular, mostly arbitrary, realisation of the stochastic process modulating the spin Hamiltonian will have to be conjured up. E.g.\ in order to realize S/T-dephasing, one could assume a stochastically modulated exchange or electron-electron dipolar interaction, but the modulation process, interaction strength, etc.\ would have to be chosen subjectively \cite{kattnig2016electron}. While the result will in second order in the perturbation only depend on the second moment of the interaction strength and the correlation time, high order contribution will be difficult to rule out in general, which renders the process idiosyncratic. In particular, for systems for which the microscopic details are unclear, this introduces an unnecessary (and possibly deceptive, if contributions of order higher than $2$ should become relevant) arbitrariness, where a description in terms of effective parameters, as contained in the Lindblad master equation might be preferred. One will furthermore have to ensure that the chosen process does not induce other processes, e.g.\ T/T-dephasing, if this is not desirable at the stage of the calculation. The advantage of this closed-system approach lays in the fact that well-established approaches for propagating closed radical pair systems subject to Haberkorn recombination can be utilized, either on the quantum level, or, if very large spin systems are to be addressed, on the semiclassical level \cite{schulten1978semiclassical, lewis2014asymmetric, manolopoulos2013improved}. In fact, the quantum propagation of the closed systems is expected to be more efficient than the MCWF approach, as it does not require the event detection of quantum jumps. Thus, the integration methodology can be optimized to the sampling of the relevant observables on a regular time grid instead, which has e.g.\ been realized efficiently by using exponential integrators based on the Arnoldi method \cite{sidje1998expokit}.

Despite the advents of the MCWF method, the integration of open spin system is a time-consuming process for all but simplistic systems of a few spins only. The fact that a large number of trajectories has to be accumulated, is however offset by the prospect that these calculations can be carried out in a parallel. In fact, as only a few state-vectors need to be stored, the individual resource requirements are modest and a massively parallel implementation on high-performance clusters is easily realized. Furthermore, when time-dependent observables, e.g. $p_1(t)$ or $p_S(t)$, are to be evaluated for the purpose of comparing with experiments, an integration accuracy of $10^{-2}$ to $10^{-3}$ might be sufficient (to realize results that are converged when plotted on the full scale of the calculation/experiment), which can often be realized with as little as $1000$ samples. While this is encouraging, the calculation of reaction yields, e.g.\ $Y_1$ and $Y_S$, does typically require more samples to converge to the required (experimental) precision. In particular, the MCWF method cannot provide a simple answer to some of the challenges posed by the spin dynamics thought to underpin magnetoreception. For these systems, long-lived coherences have been postulated, which in many cases nonetheless only give rise to small anisotropies of the reaction yields. This is particularly problematic for the prototypical [FAD$^{\bullet-}$ W$^{\bullet+}$] systems, which for realistic descriptions of the nuclear spin degrees of freedom (for the closed system dynamics; open system dynamics have hardly been addressed), give rise to but tiny relative anisotropies of often markedly less than $1 \%$ \cite{hiscock2016quantum, master_students}. In combination with the long lifetimes, this requires a substantial investment in computing time to resolve these tiny directional effects. Note, however, that much larger effects can result from the [FAD$^{\bullet-}$ Z$^{\bullet}$] or some recently introduced radical triad systems \cite{kattnig2017sensitivity, kattnig2017radical, lee2014alternative}.

We also note that the scaling behaviour of the MCWF approach, while growing weaker than that of the ME approach, is still exponential in the number of coupled spins. While algorithms for simulating spin dynamics that scale polynomially in the number of spins have become popular in the field of theoretical magnetic resonance spectroscopy, these methods do not usually provide sufficiently accurate solutions for the radical pair dynamics in weak magnetic fields. Here, we expect that the MCWF approach can fill the gap that exists for open quantum system dynamics between the toy systems that are straight-forwardly treated by the ME approach and the realm of semiclassical approaches (possibly with direct inclusion of the bath degrees of freedom) \cite{hogben2011spinach, hogben2010strategies, kuprov2007polynomially}.

The disadvantage of having to sample a large number of trajectories can sometimes emerge to be a virtue. This is the case if, e.g., the effects of random time dependent interactions are to be addressed or if a stochastic motion modulates the spin Hamiltonian, such as is the case if the radical pair undergoes mutual diffusive motion \cite{evans2016sub}. Including these otherwise difficult to accommodate aspects in the description is straight forward in the MCWF approach, e.g.\ by combining the integration of the wavefunction with a stochastic differential equations solver to accounting for these random processes. This is clearly a field where we expect to see much interest in the near future.

For systems for which the effective Hamiltonian factors due to the existence of constants of motion, e.g.\ as is the case for isotropic spin systems, large savings of CPU time can be realized by utilizing the block structure of operators. Likewise, marked improvements of the runtime and its scaling are to be expected for problems for which the effective Hamiltonian can still be diagonalized, but a description of the open state dynamics is aimed for (e.g. to include the effect of S/T-dephasing). In these cases, an efficient approach could be realized by implementing the main propagation step in between quantum jumps in the eigenbasis of the effective Hamiltonian. This is expected to provide a significant speed-up for the large class of systems of intermediate size, i.e.\ systems that are non-trivial in terms of the open system dynamics but too small to be well-described by semiclassical approaches. While we here have not explored this idea further, we expect that future application focused on actual applications will profit from this or similar tactics. In fact, here we have not considered the symmetry decomposition of the Hilbert space that could have been realized as the three protons in the methyl groups at C7 and C8 of the flavin radical anion are completely equivalent and could thus be treated in a coupled spin basis. In this sense, the presented results are representative of the worst case scenario, where the structure of the effective Hamiltonian cannot be utilized in any particular way. Actual application might profit from additional efficiency boosts within the outlined MCWF approach and exceed our current application.

Eventually, while we here do not aim for new insights into the radical pair dynamics putatively underpinning magnetoreception, we want to point out a novel observation. As illustrated in Fig.\ \ref{1}, S/T-dephasing appears to strongly attenuate the anisotropy of the magnetic field effect, i.e.\ in the presence of this relaxation process with $\gamma_{ST} = 11$ \si{\us}$^{-1}$ the orientational differences in $f_1(t)$ are essentially washed out. While obviously more detailed studies of this effect are mandated, this once again highlights the importance of focusing on the study of open system dynamics to unravel the true nature of biological magnetic field effect. Indeed, so far no credible demonstration of magnetic anisotropy has been realized for the [FAD$^{\bullet-}$ W$^{\bullet+}$] radical pair.  

\section{Conclusions}

We have shown that the quantum Monte Carlo wavefunction (MCWF) approach can be extended to the non-Lindbladian master equations relevant to the spin dynamics of radical systems. We achieved this by stipulating that when radicals recombine they are no longer active in the interactions of the system, and thus the path describing them should be terminated upon a recombination occurring which was added into the MCWF framework by the introduction of an additional "quantum jump" to describe the termination step.
We have tested this new approach against the benchmark of direct integration of the master equation with a Runge-Kutta approach, and find that the asymptotic time-complexity of the MCWF approach is $\mathcal{O}(2.4^n)$ compared to the master equation scaling of $\mathcal{O}(4.5^n)$, where $n$ is the number of protons in the system. This speed-boost allows large spin systems of up to twenty fully-interacting spins to be integrated, where previously this was computationally unfeasible. Small spin systems, on the other hand, are better treated using the traditional, direct integration of the master equation. We expect that the MCWF method will become a useful asset in the study of magnetoreception and other biological magnetic field effects that are discussed in the context of the emerging discipline of quantum biology. In this context, the complexity of the radical systems of biological relevance in terms of the large number of hyperfine coupled nuclear spins has so far precluded the study of open system dynamics for realistic scenarios. Here, we expect the delineated approach to significantly broaden the range of problems that can be routinely analysed.

\section{Acknowledgments}

We gladly acknowledge the use of the University of Exeter High-Performance Computing (HPC) facility in carrying out this work and the EPSRC (Grant EP/R021058/1) for financial support. D.\ R.\ K.\ thanks I.\ Kominis for valuable comments on an early draft of this manuscript.

\bibliography{References}

\end{document}